\documentclass{article}
\usepackage{graphics}
\usepackage{latexsym}
\usepackage{natbib}
\newcommand{\nc}{\newcommand}  
\nc{\teff}{$T_{\rm eff}$\,}  
\nc{\logg}{log\,$g$\,}  
\nc{\kms}{\,${\rm km\,s}^{-1}$\,}  
\nc{\mic}{$\xi_{\rm t}$\,}
\begin{document}

\title{Elemental abundances of metal poor carbon rich lead star: CS29497-030
\thanks{Talk given at the meeting {\it Early Galactic Chemical Evolution 
with UVES} held at ESO Garching on 29-30th November 2002} }
\maketitle
\author{ T. Sivarani$^{1}$ , P. Bonifacio$^{1}$, P. Molaro$^{1}$, R. Cayrel$^{2}$, 
M. Spite $^{2}$, F. Spite$^{2}$, B.Plez$^{3}$, J. Andersen$^{4}$, B. Barbuy$^{5}$,
T.C. Beers$^{6}$, E. Depagne$^{2}$, V.Hill$^{2}$, P. Fran\c cois$^{2}$,
B. Nordstr\"om$^{7,4}$, and F. Primas$^{8}$ } \\
\\
$^{1}$ Istituto Nazionale di Astrofisicia - Osservatorio Astronomico di Trieste,
Via Tiepolo 11, I-34131 Trieste, Italy \\
$^2$Observatoire de Paris-Meudon, GEPI,
             F-92195 Meudon Cedex, France\\
  $^3$   GRAAL, Universit\'e de Montpellier II, F-34095
  Montpellier Cedex 05, France\\
  $^4$         Astronomical Observatory, NBIfAFG, Juliane Maries Vej 30,
           DK-2100 Copenhagen, Denmark\\
$^5$     IAG, Universidade de S\~ao Paulo, Departamento de Astronomia, CP
         3386, 01060-970 S\~ao Paulo, Brazil\\
$^6$              Department of Physics \& Astronomy, Michigan State
University,
             East Lansing, MI 48824, USA\\
$^7$         Lund Observatory, Box 43, S-221 00 Lund, Sweden\\
$^8$          European Southern Observatory (ESO),
         Karl-Schwarschild-Str. 2, D-85749 Garching b. M\"unchen,
Germany\\

We present here the abundance analysis of a metal poor carbon
rich lead star, CS29497-030. High resolution and high signal to noise
spectra were obtained using the UVES spectrograph
on the 8.2m VLT-Kueyen telescope. The observations
were made as a part of the Large Programme  165.N-0276, P.I. R. Cayrel.
Abundance analysis was done using the latest version of the 
MARCS model atmospheres (Plez et. al. 1992) and the
turbospectrum spectrum synthesis code. We have derived \teff = 6650K
from  the FeI lines. Visible and infrared broad band colours using 
the Alonso et al. (1996)
calibration,  gives similar temperatures.  A \logg value of 3.5
was obtained from the ionisation equilibrium of FeI and FeII, we
remark that this gravity 
also satisfies the MgI/MgII, Ti I/Ti II and MnI/MnII  equilibria,
within errors.
The abundance analysis indicates a metallicity,
[Fe/H] = --2.7. A  large overabundance of carbon ([C/Fe]=2.7) was
found.
We have also found large enhancement in the s-process elements
and in particular  lead shows an extremely high abundance of [Pb/Fe]=3.5,
which makes this the star with the highest Pb/Fe ratio,  up to date.
The Pb/Ba ratio is found to be high ([Pb/Ba]=1.2) and the same
is true for other
second-peak s-process elements(e.g  La, Ce, Nd). 
The star is a known spectroscopic binary with a period
of 346 days (Preston \& Sneden 2000).
The abundance pattern  suggests that 
CS 29497-30  
has accreted matter from its  companion, when it was in the AGB phase. 

\vfill\eject
\begin{table}
\caption{Log of observations}
\begin{tabular}{llll}
\hline
Date & Exp. time(s) & $\lambda_c$ (nm)           & JD \\
\hline
19/10/2000  & 1800         & 396+850             & 51836.2060193  \\
06/11/2001  & 1800         & 396+573             & 52219.1752297  \\

\hline
\end{tabular}
\end{table}
      
\begin{table}
\caption{Atmospheric Parameters of CS 29497$-$030}
\begin{tabular}{lllll}
\hline
Magnitude \& colour    &       & $\sigma$ & \teff(Alonso) &$\delta$\teff$^{1}$\\
\hline
V                      & 12.65 &  0.002   &               &       \\ 
%(B-V)                 & 0.299 &  0.004   & 6525          & 20    \\
(B-V)$_{o}$            & 0.315 &  0.004   & 6525          & 20    \\
%(V-R)                 & 0.215 &  0.003   & 6690          & 24    \\
(V-R)$_{o}$            & 0.343 &  0.003   & 6690          & 24    \\
%(V-K)$_{o}$           & 0.960 &  0.037   & 6840          & 91    \\
%(J-K)$_o {2mass}$     & 0.225 &  0.054   & 6520          & 400   \\
%(J-H)$_o {2mass}$     & 0.179 &  0.067   & 6660          & 611   \\
\hline
Adopted parameters     & \teff & \logg & \mic \kms & [Fe/H]    \\
from FeI \& FeII lines & 6650  & 3.5 &  2.0        & -2.7      \\
\hline
\end{tabular}

{\it $^{1}$ Change in \teff for +$\sigma$ change in the color.\\
Alonso- Alonso et al.(1996)}
\end{table}

\begin{table}
\caption{}
\begin{center}
\begin{tabular}{llllll}
\hline
Species & Solar X/H & [X/H]  & $\sigma$ & no. lines  & comments\\
\hline
C I     & 8.52      & -0.01  &   0.24   &  14        & NLTE  not included \\
CH      &           & -0.32  &          &            &          \\
O I     &  8.83     & -1.03  &          &   3        & From the 7774 lines. \\
        &           &        &          &            & NLTE not included \\
NaI     & 6.33      & -2.18  &   0.15   &  2         & From the D lines   \\
MgI     & 7.58      & -2.16  &   0.21   &  3         &  \\
MgII    &           & -2.06  &          &            & From 4481 lines \\
Al I    &  6.47     & -3.47  &          &  1         &  \\
Si I    &  7.55     & -2.82  &          &  1         &  \\
S I     &  7.33     & -1.63  &          &  2         & weak lines \\
Ca I    &  6.36     & -2.37  &  0.17    &  9         &  \\
Ti I    &  5.02     & -2.54  &          &  1         &  \\
Ti II   &           & -2.41  &  0.28    &  19        &  \\
Cr I    &  5.67     & -2.87  &  0.12    &  6         &  \\
Cr II   &           & -2.75  &  0.07    &  4         &  \\
Mn I    &  5.39     & -2.89  &  0.28    &  3         & \\
Mn II   &           & -2.87  &  0.00    &  2         & \\
Fe I    &  7.50     & -2.77  &  0.14    &  55        & \\
Fe II   &           & -2.70  &  0.10    &  5         & \\
Co I    &  4.92     & -2.28  &          &  1         & \\
Ni I    &  6.25     & -2.91  &  0.22    &  11        & \\
Sr II   &  2.97     & -1.86  &  0.04    &  2         & \\
Y II    &  2.24     & -2.14  &          &  3         & from synthesis \\
Zr II   &  2.60     & -1.73  &          &  1         & \\
Ba II   &  2.13     & -0.43  &          &  2         & \\
La II   &  1.17     & -0.45  & 0.78     &  3         & \\
Ce II   &  1.58     & -0.42  & 0.00     &  2         & \\
Eu II   &  0.51     & -1.41  &          &            & \\
Pb I    &  1.95     &  0.75  &          &  1         & Isotopic correction and fine \\
        &           &        &          &            & structure splitting not included   \\
\hline
\end{tabular}
\end{center}
\end{table}
\begin{table}
\caption{}
\begin{tabular}{llllllll}
\hline
 Object             & Teff/logg & [Fe/H] & [C/Fe] & [Ba/Fe] &  [Pb/Fe]  & Binary & Period\\
 \hline
 CS 29526-110$^{1}$ & 6500,3.2  & -2.38  &  2.2   &    2.11 &     3.3  &     yes&\\
 CS 22898-027$^{1}$ & 6250,3.7  & -2.26  &  2.2   &    2.23 &     2.84 &     no&\\
 CS 31062-012$^{1}$ & 6250,4.5  & -2.55  &  2.1   &    1.98 &     2.4  &     --&\\
 CS 22880-074$^{1}$ & 5850,3.8  & -1.93  &  1.3   &    1.31 &     1.9  &     no&\\
 CS 31062-050$^{1}$ & 5600,3.0  & -2.31  &  2.0   &    2.30 &     2.9  &     --&\\
 CS 22942-019$^{1}$ & 5000,2.4  & -2.64  &  2.0   &    1.92 &   $\le1.6$  &     yes &   2800\\
 CS 30301-015$^{1}$ & 4750,0.8  & -2.64  &  1.6   &    1.45 &     1.7  &     --&\\
 HE0024-2523$^{2}$  & 6625,4.3  & -2.72  &  2.6   &    1.46 &     3.3  &     yes& 3.14 \\
 LP 625-44$^{1a}$    & 5500,2.8  & -2.71  &  2.1   &    2.74 &     2.6  &     yes&\\
 LP 706-7$^{1b}$     & 6000,3.8  & -2.74  &  2.15  &    2.01 &     2.28 &     no&\\
 CS 29497-030$^{3}$ & 6650,3.5  & -2.77  &  2.45  &    2.34 &     3.5  &   yes& 346\\
 HD 196944$^{1}$    & 5250,1.8  & -2.25  &  1.2   &    1.07 &     1.7  &     --&\\
 \hline
 \end{tabular}
 {\it (1) Aoki et al. 2002 (1a) Aoki et al 2002a  (1b) Aoki et al. 2001 (2) Lucatello et al. 2002 (3) Present work}
\end{table}

\begin{figure}
\rotatebox{90}{\resizebox{8cm}{15cm}{\includegraphics{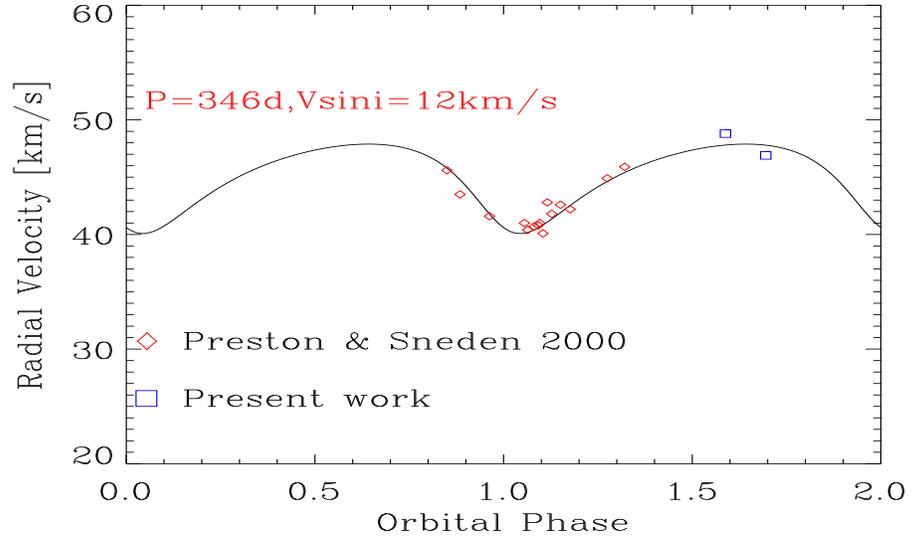}}}
\caption{The data taken from Preston \& Sneden (2000) is indicated in 
$\Diamond$ and the $\Box$ is our observations. The orbital parameters
shown are taken from Preston and Sneden (2000)}
\label{}
\end{figure}

\begin{figure}
\rotatebox{90}{\resizebox{6cm}{11cm}{\includegraphics{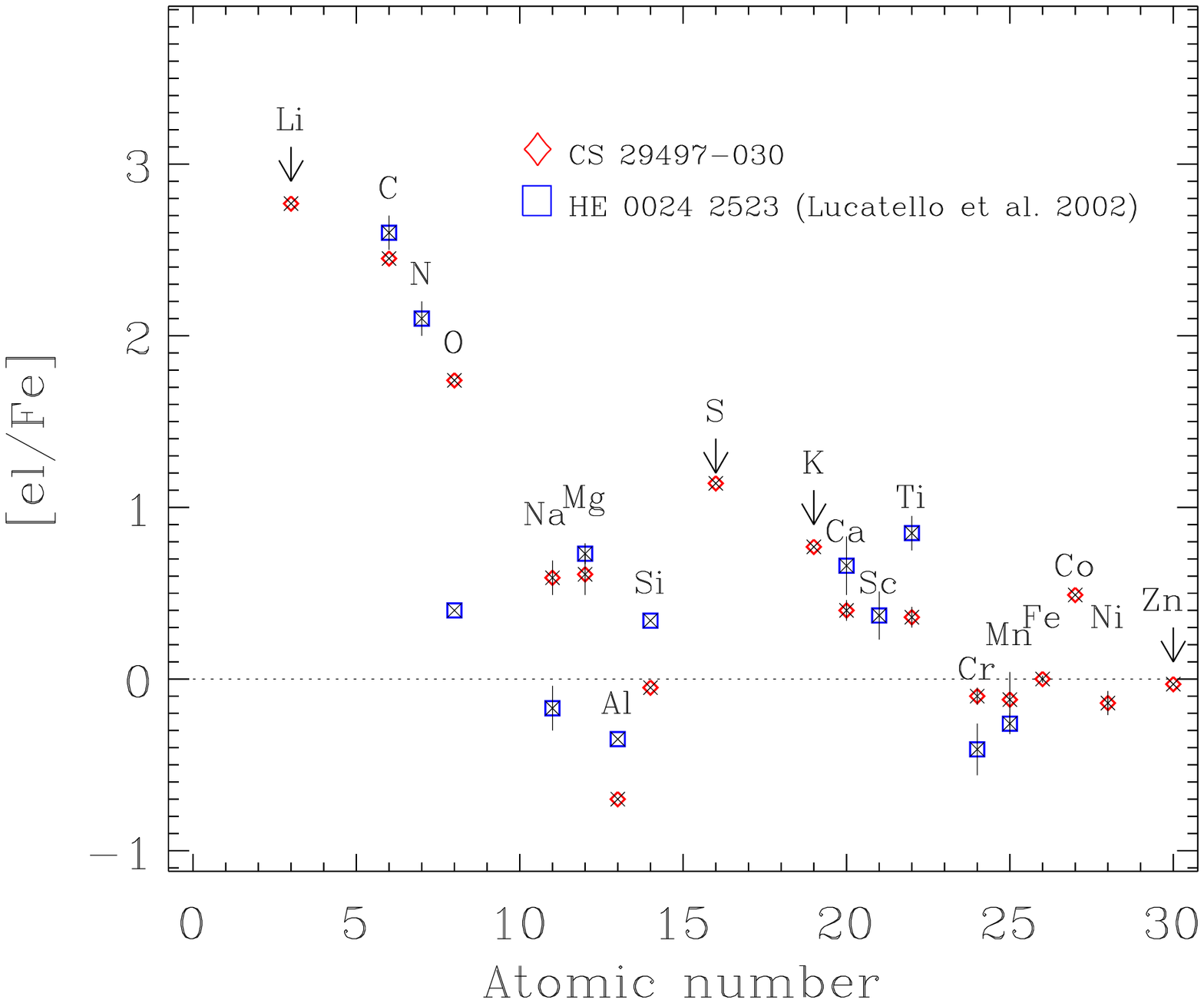}}}
\caption{Abundance pattern}
\label{}
\end{figure}

\begin{figure}
\rotatebox{90}{\resizebox{6cm}{11cm}{\includegraphics{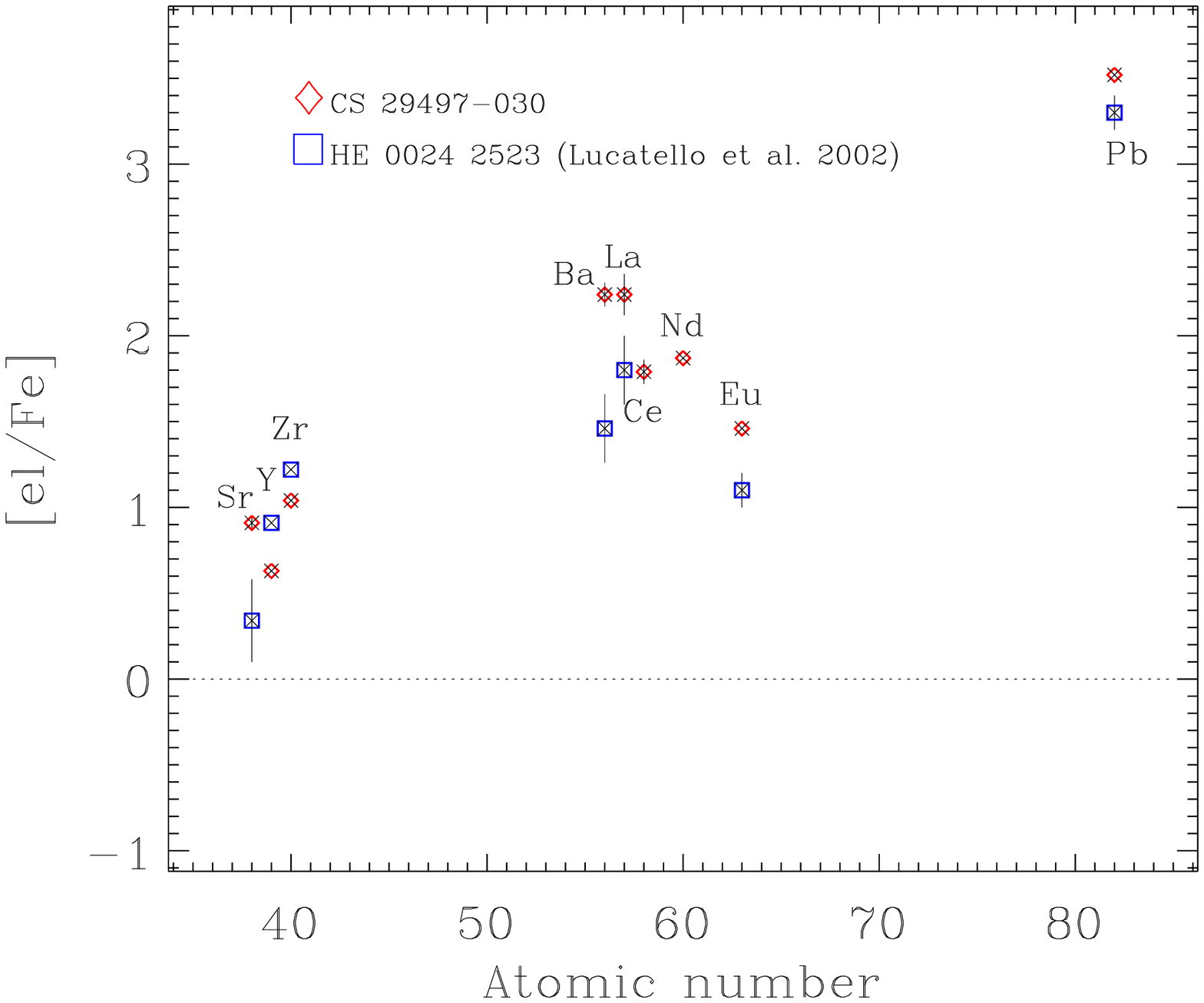}}}
\caption{Abundance pattern}
\label{}
\end{figure}

\begin{figure}
\rotatebox{90}{\resizebox{8cm}{15cm}{\includegraphics{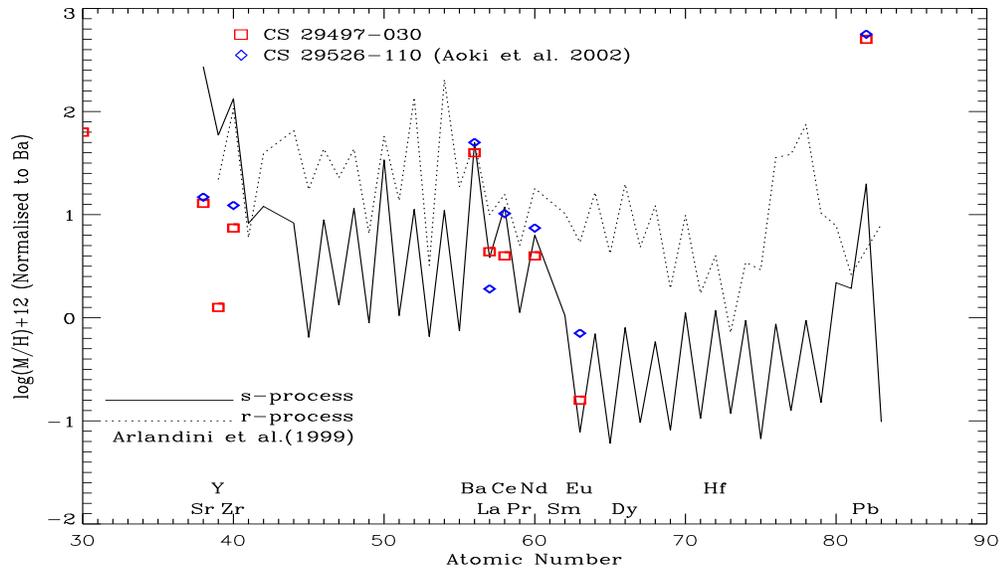}}}
\caption{The solid line indicates the main s-process component 
determined by Arlandini et al.(1999), while the dotted line indicates the 
r-process component. $\Box$ symbol indicates the abudances for CS 29497-030
and $\diamond$ represents CS 31062-050(from Aoki et al. 2002). The abundance
pattern is normalised to Ba.}
\label{}
\end{figure}

\begin{figure}
\rotatebox{90}{\resizebox{8cm}{15cm}{\includegraphics{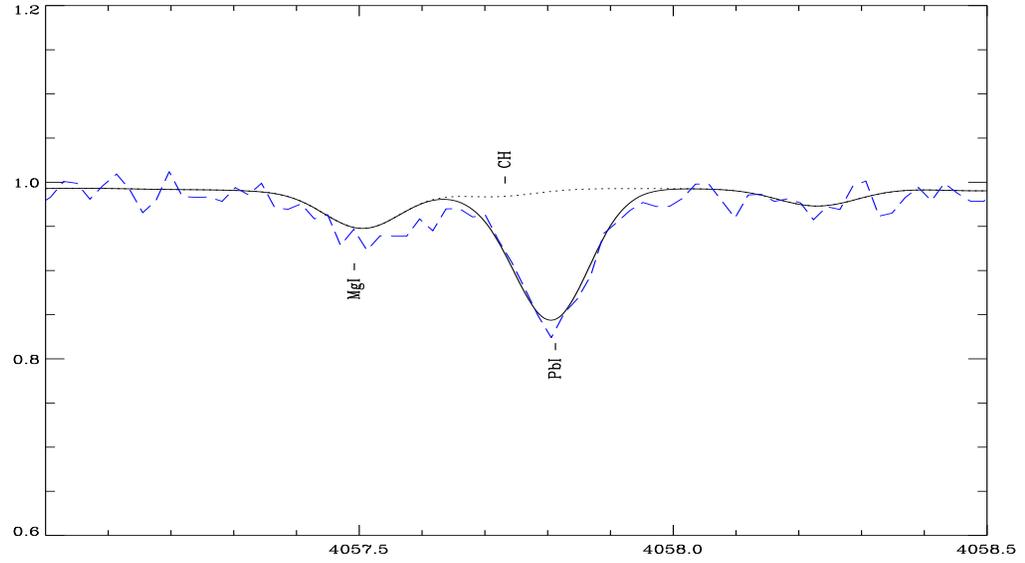}}}
\caption{The dashed line is the observed spectra and the solid line
is the synthetic spectra. The dotted line shows the contribution of
CH line blend.
}
\label{}
\end{figure}

\begin{figure}
\rotatebox{90}{\resizebox{8cm}{15cm}{\includegraphics{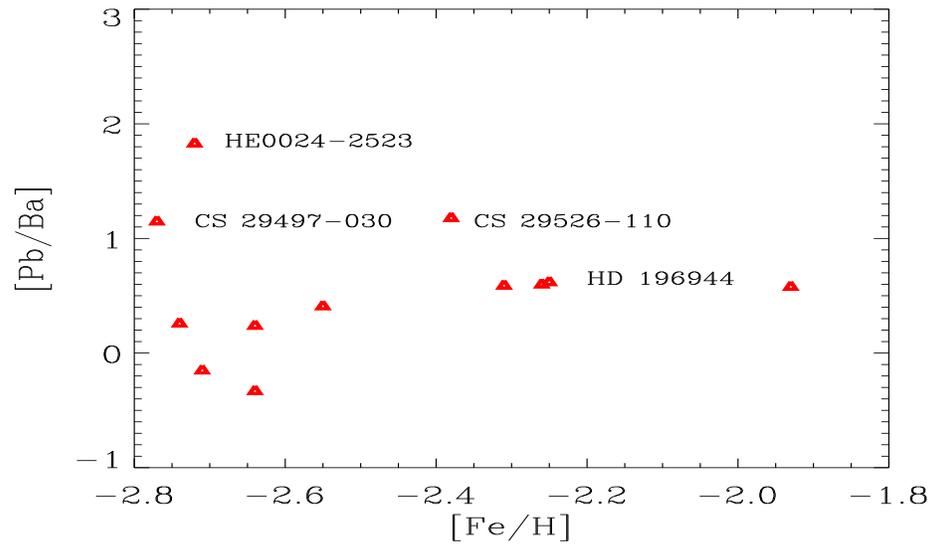}}}
\caption{The data plotted here are  from the list in table.4}
\end{figure}
\newpage
\bibliographystyle{aa}

\end{document}